\documentclass[]{INCLUDES/llncs}
\usepackage{amsmath}
\usepackage{graphicx}
\usepackage{subfig}
\usepackage{url}
\usepackage{verbatim} 
\usepackage{listings}
\lstnewenvironment{code}
    {\lstset{}%
      \csname lst@SetFirstLabel\endcsname}
    {\csname lst@SaveFirstLabel\endcsname}
\newtheorem{prop}{Proposition}
\newtheorem{df}{Definition}

\def \bscale1 {1.00}
\def \bscale {0.20}

\newcommand{\FIG}[3]{
\begin{figure}[htbp]
\centering
\scalebox{\bscale1}{\includegraphics*[bb=0pt 0pt 800pt 800pt]{./figs/#3.png}}
\caption{#2}
\label{#1}
\end{figure}
}

\begin{document}

\title{
  Ranking and Unranking of Hereditarily Finite Functions
  and
  Permutations
}

\author{Paul Tarau}
\institute{
   Department of Computer Science and Engineering\\
   University of North Texas\\
   {\em E-mail: tarau@cs.unt.edu}
}

\maketitle

\date{}

\begin{abstract}
Prolog's ability to return multiple answers on backtracking
provides an elegant mechanism to derive reversible encodings of combinatorial
objects as Natural Numbers i.e. {\em ranking} and {\em unranking} functions.
Starting from a generalization of Ackerman's encoding of Hereditarily Finite
Sets with Urelements and a novel tupling/untupling operation, we derive
encodings for Finite Functions and use them as building blocks for an executable
theory of {\em Hereditarily Finite Functions}. The more difficult problem
of {\em ranking} and {\em unranking} {\em Hereditarily Finite Permutations} is
then tackled using Lehmer codes and factoradics.

The paper is organized as a self-contained literate Prolog program  available 
at \url{http://logic.csci.unt.edu/tarau/research/2008/pHFF.zip}.

\end{abstract}

\section{Introduction}

This paper is an exploration with logic programming tools of {\em ranking} and
{\em unranking} problems on finite functions and bijections and their related
hereditarily finite universes. The practical expressiveness of logic programming
languages (in particular Prolog) are put at test in the process. The paper is part
of a larger effort to cover in a declarative programming 
paradigm, arguably more elegantly, some fundamental combinatorial generation 
algorithms along the lines of \cite{knuth06draft}.

The paper is organized
as follows: section \ref{unrank} introduces generic 
ranking/unranking functions, section \ref{ack} introduces
 Ackermann's encoding in the more general case 
when {\em urelements} are present.
Section \ref{pairings} introduces new 
tupling/untupling operations on natural numbers and uses them for encodings 
of finite functions (section \ref{fun}), resulting in encodings for 
Hereditarily Finite Functions (section \ref{hff}).
Ranking/unranking of permutations and Hereditarily Finite
Permutations as well as Lehmer codes and factoradics are covered in 
section \ref{perm}. Sections \ref{related} and \ref{concl}
discuss related work, future work and conclusions.
 
We will assume that the underlying
Prolog system supports the usual higher order 
function-style predicates {\tt
call/N}, {\tt findall/3}, {\tt maplist/N}, {\tt sumlist/2} or 
their semantic equivalents 
and a few well known library predicates, used mostly for list processing and
arithmetics. Arbitrary length integers are needed for 
some of the larger examples but 
their absence does not affect the correctness of
the code within the integer range provided by a given Prolog implementation.
Otherwise, the code in the paper, embedded in a literate programming LaTeX
file, is self contained and runs under {\em SWI-Prolog}. Note also that a few
utility predicates, not needed for following the main ideas of the paper,
are left out from the narrative and provided in the Appendix.

\section{Generic Unranking and Ranking with Higher Order Functions}
\label{unrank}
We will use, through the paper, a generic {\em multiway tree} type
distinguishing between atoms represented as (arbitrary length) integers and
subforests represented as Prolog lists.
Atoms will be
mapped to natural numbers in {\tt [0..Ulimit-1]}. 
Assuming that {\tt Ulimit} is fixed, we denote $A$ the set {\tt [0..Ulimit-1]}.
We denote $Nat$ the set of natural numbers and $T$ 
the set of trees of type $T$ with atoms in $A$.
\begin{df}
A ranking function on $T$ is a bijection $T \rightarrow Nat$.
An unranking function is a bijection $Nat \rightarrow T$.
\end{df}
{\em Ranking} functions can be traced back to 
G\"{o}del numberings \cite{Goedel:31,conf/icalp/HartmanisB74} 
associated to formulae. However, G\"{o}del numberings are typically only
injective functions, as their use in the proofs of G\"{o}del's incompleteness
theorems only requires injective mappings from well-formed formulae to
numbers. Together with their inverse {\em unranking} functions
they are also used in combinatorial and uniform random instance generation
\cite{conf/mfcs/MartinezM03,knuth06draft} algorithms.

\subsection{Unranking} As an adaptation of the {\em unfold} 
operation \cite{DBLP:journals/jfp/Hutton99,DBLP:conf/fpca/MeijerH95}, 
elements of $T$ will be mapped to natural numbers with a generic 
higher order function {\tt unrank\_} parameterized by the the natural number
{\tt Ulimit} and the transformer function {\tt F}:

\begin{code}
unrank_(Ulimit,_,N,R):-N>=0,N<Ulimit,!,R=N.
unrank_(Ulimit,F,N,R):-N>=Ulimit,
  N0 is N-Ulimit,
  call(F,N0,Ns),
  maplist(unrank_(Ulimit,F),Ns,R).
\end{code}
 A global constant provided by the predicate {\tt default\_ulimit},
will be used
through the paper to fix the default range of atoms as
 well as a default {\tt unrank} function:
Note also that we will use a syntactically more convenient {\tt DCG} 
notation,
as {\tt default\_ulimit} will act as a modifier for functional style
predicates, composed by chaining their arguments automatically with 
Prolog's {\tt DCG} transformation:
\begin{code}
default_ulimit(0)-->[].

unrank(F)-->default_ulimit(Ulimit),unrank_(Ulimit,F).
\end{code}

\subsection{Ranking} Similarly, as an adaptation of {\em fold}, generic inverse 
mappings {\tt rank\_(Ulimit,G)} and {\tt rank} from $T$ to $Nat$ are defined
as:
\begin{code}
rank_(Ulimit,_,N,R):-integer(N),N>=0,N<Ulimit,!,R=N.
rank_(Ulimit,G,Ts,R):-maplist(rank_(Ulimit,G),Ts,T),call(G,T,R0),R is R0+Ulimit.

rank(G)-->default_ulimit(Ulimit),rank_(Ulimit,G).  
\end{code}
Note that the guard in the second definition simply states 
correctness constraints ensuring 
that atoms belong to the same set $A$ for {\tt rank\_} and {\tt unrank\_}. 
This ensures that the following holds:

\begin{prop}
If the transformer function $F:Nat \rightarrow [Nat]$ is a bijection with
inverse G, such that $n \geq ulimit \wedge F(n)=[n_0,...n_i,...n_k]
\Rightarrow n_i<n$, then {\tt unrank} is a bijection from $Nat$ to $T$, 
with inverse {\tt rank} and the recursive computations 
of both functions terminate in a finite number of steps.

\noindent {\em Proof:} by induction on the structure of $Nat$ and $T$, using the
fact that {\tt maplist} preserves bijections.
\end{prop}

\section{Hereditarily Finite Sets and Ackermann's Encoding} \label{ack}

The Universe of Hereditarily Finite Sets is best known as a model of the 
Zermelo-Fraenkel Set theory with the Axiom of Infinity  replaced by its 
negation \cite{finitemath,DBLP:journals/jct/MeirMM83}. In a Logic Programming
framework, it has been used for reasoning with sets, set constraints,
hypersets and bisimulations 
\cite{dovier00comparing,DBLP:journals/tplp/PiazzaP04}.

The Universe of Hereditarily Finite Sets is built from the empty set 
(or a set of {\em Urelements}) 
by successively applying powerset and set union operations.

Ackermann's encoding \cite{ackencoding,abian78,kaye07} is a bijection that
maps Hereditarily Finite Sets ($HFS$) to Natural Numbers ($Nat$) as follows:

\vskip 0.5cm
$f(x)$ = {\tt if} $x=\{\}$ {\tt then} $0$ {\tt else} $\sum_{a \in x}2^{f(a)}$
\vskip 0.5cm

Assuming $HFS$ extended with {\em Urelements} (atomic objects not having any
elements) our generic tree representation can be used for Hereditarily Finite Sets.

Ackermann's encoding can be seen as the recursive application of 
a bijection {\tt set2nat} from finite subsets of $Nat$ to $Nat$, that associates to 
a set of (distinct!) natural numbers a (unique!) natural number. 
\begin{code} 
set2nat(Xs,N):-set2nat(Xs,0,N).

set2nat([],R,R).
set2nat([X|Xs],R1,Rn):-R2 is R1+(1<<X),set2nat(Xs,R2,Rn).
\end{code}
With this representation, Ackermann's encoding from $HFS$ to $Nat$ {\tt hfs2nat} 
can be expressed in terms of our generic {\tt rank} function as:
\begin{code}
hfs2nat-->default_ulimit(Ulimit),hfs2nat_(Ulimit).

hfs2nat_(Ulimit)-->rank_(Ulimit,set2nat).
\end{code}
where the constant provided by {\tt default\_ulimit} controls the
segment {\tt [0..Ulimit-1]} of $Nat$ to be mapped to urelements. 
The default value {\tt 0} defines ``pure" sets, 
all built from the empty set only.

To obtain the inverse of the Ackerman encoding, we first define the 
inverse {\tt nat2set} of the bijection {\tt set2nat}. It decomposes a natural
number $N$ into a list of exponents of 2 (seen as bit positions equaling 1 
in $N$'s bitstring representation, in increasing order).
\begin{code}
nat2set(N,Xs):-nat2elements(N,Xs,0).

nat2elements(0,[],_K).
nat2elements(N,NewEs,K1):-N>0,
  B is /\(N,1),N1 is N>>1,K2 is K1+1,add_el(B,K1,Es,NewEs),
  nat2elements(N1,Es,K2).

add_el(0,_,Es,Es).
add_el(1,K,Es,[K|Es]).
\end{code}
The inverse of the  Ackermann encoding, with urelements in {\tt [0..Ulimit-1]}
and {\tt Ulimit} mapped to {\tt []} follows:
\begin{code}
nat2hfs_(Ulimit)-->unrank_(Ulimit,nat2set).

nat2hfs-->default_ulimit(Ulimit),nat2hfs_(Ulimit).
\end{code}

Using an equivalent functional notation, the following proposition summarizes
the results in this subsection:
\begin{prop}
Given id = $\lambda x.x$, the following function equivalences hold:

\begin{equation}
nat2set \circ set2nat \equiv id \equiv set2nat \circ nat2set
\end{equation}

\begin{equation}
nat2hfs \circ hfs2nat \equiv id \equiv hfs2nat \circ nat2hfs
\end{equation}
\end{prop}

\section{Pairing Functions and Tuple Encodings} \label{pairings}

{\em Pairings} are bijective functions $Nat \times Nat \rightarrow Nat$.  
We refer to  \cite{DBLP:journals/tcs/CegielskiR01} for a typical use 
in the foundations of mathematics and to \cite{DBLP:conf/ipps/Rosenberg02a} 
for an extensive study of various pairing functions and their computational properties. 

\subsection{The Pepis-Kalmar-Robinson Pairing Function}

The predicates {\tt pepis\_pair/3} and {\tt pepis\_unpair/3} are derived from
the function {\bf pepis\_J} and its left and right unpairing companions 
{\bf pepis\_K} and {\bf pepis\_L} that have been used, by Pepis, 
Kalmar and Robinson 
in some fundamental work on recursion
theory, decidability and Hilbert's Tenth Problem 
in \cite{pepis,kalmar1,robinson67}:
\begin{code}
pepis_pair(X,Y,Z):-pepis_J(X,Y,Z).

pepis_unpair(Z,X,Y):-pepis_K(Z,X),pepis_L(Z,Y).
 
pepis_J(X,Y, Z):-Z is ((1<<X)*((Y<<1)+1))-1.
pepis_K(Z, X):-Z1 is Z+1,two_s(Z1,X).
pepis_L(Z, Y):-Z1 is Z+1,no_two_s(Z1,N),Y is (N-1)>>1. 

two_s(N,R):-even(N),!,H is N>>1,two_s(H,T),R is T+1.
two_s(_,0).

no_two_s(N,R):-two_s(N,T),R is N // (1<<T).

even(X):- 0 =:= /\(1,X).
\end{code}
This pairing function given by the formula

\begin{equation}
f(x,y)=2^x*(2*y+1)-1
\end{equation}

\noindent is asymmetrically growing, faster on the
first argument. It works as follows:
\begin{verbatim}
?- pepis_pair(1,10,R).
R = 41.

?- pepis_pair(10,1,R).
R = 3071.

?- findall(R,(between(0,3,A),between(0,3,B),pepis_pair(A,B,R)),Rs).
Rs=[0, 2, 4, 6, 1, 5, 9, 13, 3, 11, 19, 27, 7, 23, 39, 55]
\end{verbatim}

\subsection{Tuple Encodings} \label{tuple}
We will now generalize pairing functions to $k$-tuples and then we will 
derive an encoding for finite functions. 

The function {\tt to\_tuple:} $Nat \rightarrow Nat^k$ converts a natural 
number to a $k$-tuple by splitting its bit representation into $k$ groups, 
from which the $k$ members in the tuple are finally rebuilt. This operation 
can be seen as a transposition of a bit matrix obtained by expanding 
the number in base $2^k$:
\begin{code}  
to_tuple(K,N, Ns):-
  Base is 1<<K,to_base(Base,N,Ds),maplist(to_maxbits(K),Ds,Bss),
  mtranspose(Bss,Xss),
  maplist(from_rbits,Xss,Ns). 
\end{code}
To convert a $k$-tuple back to a natural number we will merge their 
bits, $k$ at a time. This operation uses the transposition of a bit 
matrix obtained from the tuple, seen as a number in base $2^k$, 
with help from bit crunching functions given in Appendix:
\begin{code}
from_tuple(Nss,R):-
  max_bitcount(Nss,L),length(Nss,K),maplist(to_maxbits(L),Nss,Mss),
  mtranspose(Mss,Tss),
  maplist(from_rbits,Tss,Ts),Base is 1<<K,from_base(Base,Ts,R).
\end{code}
The following example shows the decoding of {\tt 42}, its decomposition 
in bits (right to left), the formation of a $3$-tuple and the encoding 
back to {\tt 42}.
\begin{verbatim}
?- to_tuple(3,42,T),to_rbits(2,Bs2),to_rbits(1,Bs1),from_tuple(T,N).
T = [2, 1, 2],
Bs2 = [0, 1],
Bs1 = [1],
N = 42
\end{verbatim}
Note that one can now define pairing functions as instances of the tupling
functions:
\begin{code}
to_pair(N,A,B):-to_tuple(2,N,[A,B]).

from_pair(X,Y,Z):-from_tuple([X,Y],Z).
\end{code}
One can observe that {\tt to\_pair} and {\tt from\_pair} are the same as the
functions defined in Steven Pigeon's PhD thesis 
on Data Compression \cite{pigeon}, page 114).

\section{Encoding Finite Functions} \label{fun}

As finite sets can be put in a bijection with an initial segment 
of $Nat$, we can narrow down the concept of finite function as follows:
\begin{df}
A {\tt finite function} is a function defined from an initial 
segment of $Nat$ to $Nat$.
\end{df}

This definition implies that a finite function can be seen as an array or 
a list of natural numbers except that we do not limit the size of 
the representation of its values.

\subsection{Encoding Finite Functions as Tuples}

We can now encode and decode a finite function from $[0..K-1]$ to $Nat$ 
(seen as the list of its values), as a natural number:
\begin{code}
ftuple2nat([],0).
ftuple2nat(Ns, N):-Ns=[_|_],
  length(Ns,K),K1 is K-1, 
  from_tuple(Ns,T),pepis_pair(K1,T, N).

nat2ftuple(0,[]).  
nat2ftuple(N,Ns):-N>0,
  pepis_unpair(N,K,F),K1 is K+1,
  to_tuple(K1,F,Ns).
\end{code}
As the length of the tuple, {\tt K}, is usually smaller than the number 
obtained by merging the bits of the {\tt K}-tuple, we have picked the 
Pepis pairing function, exponential in its first argument and linear 
in its second, to embed the length of the tuple needed for the decoding. 
The encoding/decoding works as follows:
\begin{verbatim}
?- ftuple2nat([1,0,2,1,3],N).
N = 21295
?- nat2ftuple(21295,T).
T=[1,0,2,1,3]
?-ints_from(0,15,Is),maplist(nat2ftuple,Is,Ts).
Ts=[[0],[0,0],[1],[0,0,0],[2],[1,0],[3],
    [0,0,0,0],[4],[0,1],[5],[1,0,0],[6],
    [1,1],[7],[0,0,0,0,0]]
\end{verbatim}
Note that
\begin{code}
nat(0).
nat(N):-nat(N1),N is N1+1.

iterative_fun_generator(F):-nat(N),nat2ftuple(N,F).
\end{code}
provides an iterative generator for the stream of finite functions.

\subsection{Deriving Encodings of Finite Functions from Ackermann's Encoding}

Given that a finite set with n elements can be put in a bijection 
with $[0..N-1]$, a finite functions $f : [0..n-1] \rightarrow Nat$ 
can be represented as the list $[f(0)...f(n-1)]$. Such a list has 
however repeated elements. So how can we turn it into a set with 
distinct elements, bijectively?

The following two predicates provide the answer.

First, we just sum up the list of the values of the function, 
resulting in a monotonically growing sequence (provided that we first 
increment every number by 1 to ensure that 0 values do not break monotonicity).
\begin{code}
fun2set([],[]).
fun2set([A|As],Xs):-findall(X,prefix_sum(A,As,X),Xs).

prefix_sum(A,As,R):-append(Ps,_,As),length(Ps,L),
  sumlist(Ps,S),R is A+S+L.
\end{code}
The inverse of {\tt fun2set} reverting back from a set of distinct values 
collects the increments from a term to the next (and ignores the last one):
\begin{code}
set2fun([],[]).
set2fun([X|Xs],[X|Fs]):-set2fun(Xs,X,Fs).

set2fun([],_,[]).
set2fun([X|Xs],Y,[A|As]):-A is (X-Y)-1,set2fun(Xs,X,As).
\end{code}

\begin{prop}
The following function equivalences hold:

\begin{equation}
fun2set \circ set2fun \equiv id \equiv set2fun \circ fun2set
\end{equation}
\end{prop}

\noindent The mapping and its inverse work as follows:
\begin{verbatim}
?- fun2set([1,0,2,1,2],Set),set2fun(Set,Fun).
Set = [1, 2, 5, 7, 10],
Fun = [1, 0, 2, 1, 2].
\end{verbatim}

By combining this bijection with Ackermann's encoding's basic step {\tt
set2nat} and its inverse {\tt nat2set}, we obtain an encoding 
from finite functions to $Nat$ as follows (with DCG notation used to express
function composition):
\begin{code}  
nat2fun --> nat2set,set2fun.

fun2nat --> fun2set,set2nat.
\end{code}
\begin{verbatim}
?- nat2fun(2008,F),fun2nat(F,N).
F = [3, 0, 1, 0, 0, 0, 0], N = 2008
\end{verbatim}

\begin{prop}
The following function equivalences hold:

\begin{equation}
nat2fun \circ fun2nat \equiv id \equiv fun2nat \circ nat2fun
\end{equation}
\end{prop}

One can see that this encoding ignores {\tt 0}s in the binary representation 
of a number, while counting {\tt 1} sequences as increments. 
Alternatively, {\em Run Length Encoding} of binary sequences
\cite{conf/cpm/MakinenN05} encodes {\tt 0}s and {\tt 1}s symmetrically, 
by counting the numbers 
of {\tt 1}s and {\tt 0}s. This encoding is reversible, 
given that {\tt 1}s and {\tt 0}s alternate, 
and that the most significant digit is always {\tt 1}:
\begin{code}
bits2rle([],[]):-!.
bits2rle([_],[0]):-!.
bits2rle([X,Y|Xs],Rs):-X==Y,!,bits2rle([Y|Xs],[C|Cs]),C1 is C+1,Rs=[C1|Cs].
bits2rle([_|Xs],[0|Rs]):-bits2rle(Xs,Rs).

rle2bits([],[]).
rle2bits([N|Ns],NewBs):-rle2bits(Ns,Xs),
  ( []==Xs->B is 1
  ; Xs=[X1|_],B is 1-X1
  ),
  N1 is N+1,ndup(N1,B,Bs),append(Bs,Xs,NewBs).
\end{code}
By composing {\tt bits2rle} and  {\tt rle2bits} with converters to/from
bitlists, we obtain the bijection $nat2rle:Nat \rightarrow [Nat]$ 
and its inverse $rle2nat:[Nat] \rightarrow Nat$
\begin{code}
nat2rle --> to_rbits0,bits2rle.
rle2nat --> rle2bits,from_rbits .

to_rbits0(0,[]).
to_rbits0(N,R):-N>0,to_rbits(N,R).
\end{code}

\begin{prop}
The following function equivalences hold:

\begin{equation}
nat2rle \circ rle2nat \equiv id \equiv rle2nat \circ nat2rle
\end{equation}
\end{prop}

\section{Encodings for ``Hereditarily Finite Functions''} \label{hff}
One can now build a theory of ``Hereditarily Finite Functions" 
($HFF$) centered around using a transformer 
like {\tt nat2ftuple, nat2fun, nat2rle} and {\tt ftuple2nat, fun2nat, rle2nat} 
in way similar to the use of {\tt nat2set} and {\tt set2nat} for $HFS$, 
where the empty function (denoted {\tt []}) replaces the 
empty set as the quintessential ``urfunction''. 
Similarly to Urelements in the $HFS$ theory, ``urfunctions'' 
(considered here as atomic values) can be introduced as constant functions 
parameterized to belong to $[0..Ulimit-1]$.

By using the generic {\tt unrank\_} and {\tt rank} predicates defined 
in section \ref{unrank} we can extend the bijections defined  in this 
section to encodings of Hereditarily Finite Functions.
By instantiating the transformer function in {\tt unrank\_} to {\tt nat2ftuple}, 
{\tt nat2fun} and {\tt nat2rle} we obtain (with DCG notation expressing
composition of functional predicates):
\begin{code}
nat2hff --> default_ulimit(D),nat2hff_(D).   
nat2hff1 --> default_ulimit(D),nat2hff1_(D).
nat2hff2  --> default_ulimit(D),nat2hff2_(D).

nat2hff_(Ulimit) --> unrank_(Ulimit,nat2fun).
nat2hff1_(Ulimit) --> unrank_(Ulimit,nat2ftuple).
nat2hff2_(Ulimit) --> unrank_(Ulimit,nat2rle).
\end{code}

By instantiating the transformer function in {\tt rank} we obtain:
\begin{code}
hff2nat --> rank(fun2nat).
hff2nat1 --> rank(ftuple2nat).
hff2nat2 --> rank(rle2nat).
\end{code}

The following examples show that {\tt nat2hff}, {\tt nat2hff1} 
and {\tt nat2hff2} are indeed bijections, and that the resulting 
$HFF$-trees are typically more compact than the $HFS$-tree 
associated to the same natural number. 
\begin{verbatim}
?- nat2hff(42,H),hff2nat(H,N).
H = [[[]], [[]], [[]]],
N = 42

?- nat2hff1(42,H),hff2nat1(H,N).
H = [[[[], [], []], []]],
N = 42

?- nat2hff2(42,H),hff2nat2(H,N).
H = [[], [], [], [], [], []],
N = 42
\end{verbatim}
\noindent Note that
\begin{verbatim}
?-nat(N),nat2hff(N,HFF).
?-nat(N),nat2hff1(N,HFF).
?-nat(N),nat2hff2(N,HFF). 
\end{verbatim}
provide iterative generators for the (recursively enumerable!) 
stream of hereditarily finite functions. 

The resulting HFF with urfunctions (seen as digits) can also 
be used as generalized {\em numeral systems} with possible applications 
to building arbitrary length integer implementations.
\begin{verbatim}
?- nat2hff_(10,1234567890,HFF).
[3, 2, 0, 1, 7, 0, 1, 2, 0, 2, 2]
\end{verbatim}

\begin{prop}
The following function equivalences hold:

\begin{equation}
nat2hff1 \circ hff2nat1 \equiv id \equiv hff2nat1 \circ nat2hff1
\end{equation}

\begin{equation}
nat2hff \circ hff2nat \equiv id \equiv hff2nat \circ nat2hff
\end{equation}
\end{prop}

\section{Encoding Finite Bijections} \label{perm}
To obtain an encoding for finite bijections (permutations) 
we will first review a ranking/unranking mechanism for permutations that
involves an unconventional numeric representation, {\em factoradics}.

\subsection{The Factoradic Numeral System}
The factoradic numeral system \cite{knuth_art_1997-1} replaces digits
multiplied by power of a base $N$ with digits that multiply successive values
of the factorial of $N$. In the increasing order variant {\tt fr} the first
digit $d_0$ is 0, the second is $d_1 \in \{0,1\}$ and the $N$-th is $d_N \in
[0..N-1]$. The left-to-right, decreasing order variant {\tt fl} 
is obtained by reversing the digits of {\tt fr}.
\begin{verbatim}
?- fr(42,R),rf(R,N).
R = [0, 0, 0, 3, 1],
N = 42

?- fl(42,R),lf(R,N).
R = [1, 3, 0, 0, 0],
N = 42
\end{verbatim}
\noindent The Prolog predicate {\tt fr} handles the special case for $0$ and
calls {\tt fr1} which recurses and divides with increasing values of N
while collecting digits with {\tt mod}:
\begin{code}
% factoradics of N, right to left
fr(0,[0]).
fr(N,R):-N>0,fr1(1,N,R).
   
fr1(_,0,[]).
fr1(J,K,[KMJ|Rs]):-K>0,KMJ is K mod J,J1 is J+1,KDJ is K // J,
  fr1(J1,KDJ,Rs).
\end{code}
The reverse {\tt fl}, is obtained as follows:
\begin{code}
fl(N,Ds):-fr(N,Rs),reverse(Rs,Ds).
\end{code}
The predicate {\tt lf} (inverse of {\tt fl}) converts back to decimals by
summing up results while computing the factorial progressively:
\begin{code}
lf(Ls,S):-length(Ls,K),K1 is K-1,lf(K1,_,S,Ls,[]).

% from list of digits of factoradics, back to decimals
lf(0,1,0)-->[0].
lf(K,N,S)-->[D],{K>0,K1 is K-1},lf(K1,N1,S1),{N is K*N1,S is S1+D*N}.
\end{code}
Finally, {\tt rf}, the inverse of {\tt fr} is obtained by reversing {\tt fl}.
\begin{code}
rf(Ls,S):-reverse(Ls,Rs),lf(Rs,S).
\end{code}

\subsection{Ranking and unranking permutations of given size with Lehmer codes
and factoradics} 
The Lehmer code of a permutation $f$ is defined
as the number of indices j such that $1 \leq j < i$ and $f(j)<f(i)$
 \cite{DBLP:journals/dmtcs/MantaciR01}.
 \begin{prop}
 The Lehmer code of a permutation determines the permutation uniquely.
 \end{prop} 
The predicate {\tt perm2nth} computes a {\tt rank} 
for a permutation {\tt Ps} of {\tt Size>0}. 
It starts by first computing its Lehmer code {\tt Ls} with 
{\tt perm\_lehmer}. Then  it associates a unique natural 
number {\tt N} to {\tt Ls}, 
by converting it with the predicate {\tt lf} 
from factoradics to decimals. 
Note that the Lehmer code {\tt Ls} is used as the list of digits
in the factoradic representation.
\begin{code}
perm2nth(Ps,Size,N):-
  length(Ps,Size),Last is Size-1,
  ints_from(0,Last,Is),
  perm_lehmer(Is,Ps,Ls),
  lf(Ls,N).
\end{code}
The generation of the Lehmer code is surprisingly
simple and elegant in Prolog. We just instrument the
usual backtracking predicate generating a permutation
to remember the choices it makes, 
in the auxiliary predicate {\tt select\_and\_remember}!
\begin{code}
% associates Lehmer code to a permutation 
perm_lehmer([],[],[]).
perm_lehmer(Xs,[X|Zs],[K|Ks]):-
  select_and_remember(X,Xs,Ys,0,K),
  perm_lehmer(Ys,Zs,Ks).

% remembers selections - for Lehmer code
select_and_remember(X,[X|Xs],Xs,K,K).
select_and_remember(X,[Y|Xs],[Y|Ys],K1,K3):-K2 is K1+1,
  select_and_remember(X,Xs,Ys,K2,K3).
\end{code}  

The predicate {\tt nat2perm} provides the matching {\em unranking}
operation associating a permutation {\tt Ps} to a given {\tt Size>0} 
and a natural number {\tt N}.
\begin{code}
nth2perm(Size,N, Ps):-
  fl(N,Ls),length(Ls,L),
  K is Size-L,Last is Size-1,ints_from(0,Last,Is),
  zeros(K,Zs),append(Zs,Ls,LehmerCode),
  perm_lehmer(Is,Ps,LehmerCode).
\end{code}
Note also that {\tt perm\_lehmer} is used (reversibly!) this time
to reconstruct the permutation {\tt Ps} from its Lehmer code.
The Lehmer code is computed from the permutation's 
factoradic representation obtained by converting {\tt N} to {\tt Ls} 
and then padding it with {\tt 0}'s.
One can try out this bijective mapping as follows:
\begin{verbatim}
?- nth2perm(5,42,Ps),perm2nth(Ps,Length,Nth).
Ps = [1, 4, 0, 2, 3],
Length = 5,
Nth = 42

?- nth2perm(8,2008,Ps),perm2nth(Ps,Length,Nth).
Ps = [0, 3, 6, 5, 4, 7, 1, 2],
Length = 8,
Nth = 2008
\end{verbatim}

\subsection{A bijective mapping from permutations to $Nat$}
One more step is needed to to extend the mapping between permutations of a
given length to a bijective mapping from/to $Nat$: we will have to ``shift
towards infinity'' the starting point of each new bloc of permutations in $Nat$
as permutations of larger and larger sizes are enumerated.

First, we need to know by how much - so we compute the sum of
all factorials up to $N!$.
\begin{code}
% fast computation of the sum of all factorials up to N!
sf(0,0).
sf(N,R1):-N>0,N1 is N-1,ndup(N1,1,Ds),rf([0|Ds],R),R1 is R+1.
\end{code}
This is done by noticing that the factoradic representation of
[0,1,1,..] does just that. The stream of all such sums can now
be generated as usual:
\begin{code}
sf(S):-nat(N),sf(N,S).
\end{code}
What we are really interested into, is decomposing {\tt N} into
the distance to the
last sum of factorials smaller than {\tt N}, {\tt N\_M}
and its index in the sum, {\tt K}.
\begin{code}
to_sf(N, K,N_M):-nat(X),sf(X,S),S>N,!,K is X-1,sf(K,M),N_M is N-M.
\end{code}
{\em Unranking} of an arbitrary permutation is now easy - the index {\tt K}
determines the size of the permutation and {\tt N\_M} determines
the rank. Together they select the right permutation with {\tt nth2perm}.
\begin{code}
nat2perm(0,[]).
nat2perm(N,Ps):-to_sf(N, K,N_M),nth2perm(K,N_M,Ps).
\end{code}
{\em Ranking} of a permutation is even easier: we first compute
its {\tt Size} and its rank {\tt Nth}, then we shift the rank by 
the sum of all factorials up to {\tt Size}, enumerating the
ranks previously assigned.
\begin{code}
perm2nat([],0).
perm2nat(Ps,N) :-perm2nth(Ps, Size,Nth),sf(Size,S),N is S+Nth.
\end{code}
\begin{verbatim}
?- nat2perm(2008,Ps),perm2nat(Ps,N).
Ps = [1, 4, 3, 2, 0, 5, 6],
N = 2008
\end{verbatim}
As finite bijections are faithfully represented by permutations,
this construction provides a bijection from $Nat$ to 
the set of Finite Bijections.
\begin{prop}
The following function equivalences hold:
\begin{equation}
nat2perm \circ perm2nat \equiv id \equiv perm2nat \circ nat2perm
\end{equation}
\end{prop}

\subsection{Hereditarily Finite Permutations}

By using the generic {\tt unrank\_} and {\tt rank} predicates defined 
in section \ref{unrank} we can extend the {\tt nat2perm} and {\tt perm2nat} to
encodings of Hereditarily Finite Permutations ($HFP$).
\begin{code}
nat2hfp --> default_ulimit(D),nat2hfp_(D).   
nat2hfp_(Ulimit) --> unrank_(Ulimit,nat2perm).
hfp2nat --> rank(perm2nat).
\end{code}
The encoding works  as follows:
\begin{verbatim}
?- nat2hfp(42,H),hfp2nat(H,N),write(H),nl.
H = [[], [[], [[]]], [[[]], []], [[]], [[], [[]], [[], [[]]]]],
N = 42
\end{verbatim}
\begin{prop}
The following function equivalences hold:
\begin{equation}
nat2hfp \circ hfp2nat \equiv id \equiv hfp2nat \circ nat2hfp
\end{equation}
\end{prop}

\section{Related work} \label{related}
Natural Number encodings of Hereditarily Finite Sets have 
triggered the interest of researchers in fields ranging from 
Axiomatic Set Theory and Foundations of Logic to 
Complexity Theory and Combinatorics
\cite{finitemath,kaye07,DBLP:journals/mlq/Kirby07,abian78,DBLP:journals/jsyml/Booth90,DBLP:journals/jct/MeirMM83,DBLP:conf/foiks/LeontjevS00,DBLP:journals/tcs/Sazonov93,avigad97}. 
Computational and Data Representation aspects of Finite Set Theory 
have been described in logic programming and theorem proving contexts 
in \cite{dovier00comparing,DBLP:journals/tplp/PiazzaP04,DBLP:conf/types/Paulson94}. 
Pairing functions have been used work on decision problems as early 
as \cite{pepis,kalmar1,robinson50,robinsons68b}. The tuple functions
we have used to encode finite functions are new. 
While finite functions have been used extensively in various branches of mathematics 
and computer science, we have not seen any formalization of hereditarily 
Finite Functions or Hereditarily Finite Bijections as such in the literature.

\section{Conclusion and Future Work} \label{concl}

We have shown the expressiveness of logic programming as a
metalanguage for executable mathematics, by describing
natural number encodings, tupling/untupling and ranking/unranking functions
for finite sets, functions and permuations and by extending them in a
generic way to Hereditarily Finite Sets, Hereditarily Finite Functions
and Hereditarily Finite Permutations.

In a Genetic Programming context \cite{koza92,poli08}, 
the bijections between bitvectors/natural numbers 
on one side, and trees/graphs representing HFSs, HFFs, HPPs on the other side, 
suggest exploring the mapping and its action on various transformations 
as a phenotype-genotype connection.

We also foresee interesting applications in cryptography and steganography. 
For instance, in the case of the permutation related encodings -  something as
simple as the order of the cities visited or the order of names 
on a greetings card, seen as a permutation with respect to their 
alphabetic order, can provide a steganographic encoding/decoding of a secret
message by using predicates like {\tt nat2perm} and {\tt perm2nat}.

Last but not least, the use of a logic programming language to express
in a generic way some fairly intricate combinatorial algorithms
predicts an interesting new application area.

\bibliographystyle{plain}

\appendix
\section{Appendix}
To make the code in the paper fully self contained, 
we list here some auxiliary functions.

\paragraph{Integer list operations} 
These are some simple utility predicates:
\begin{code}
% generates integers From..To
ints_from(From,To,Is):-findall(I,between(From,To,I),Is).

% replicates X, N times
ndup(0, _,[]).
ndup(N,X,[X|Xs]):-N>0,N1 is N-1,ndup(N1,X,Xs).
  
zeros(N,Zs):-ndup(N,0,Zs).
\end{code}

\paragraph{Matrix Transposition}
This code transposes a matrix represented as list of lists.
\begin{code}
mtranspose([],[]):-!.
mtranspose([Xs],Css):-!,to_columns(Xs,Css).
mtranspose([Xs|Xss],Css2):-!,
  mtranspose(Xss,Css1),
  to_columns(Xs,Css1,Css2).
	
to_columns([], []).
to_columns([X|Xs],[[X]|Zs]):-to_columns(Xs,Zs).

to_columns([],Css,Css).
to_columns([X|Xs],[Cs|Css1],[[X|Cs]|Css2]) :- to_columns(Xs,Css1,Css2).    
\end{code}

\paragraph{Bit crunching functions} 
The following functions implement conversion operations
between bitlists and numbers.
Note that our bitlists represent binary numbers by 
selecting exponents of 2 in 
increasing order (i.e. ``right to left"). 
\begin{code}
% conversion to list of digits in given base
to_base(Base,N,Bs):-to_base(N,Base,0,Bs).

to_base(N,R,_K,Bs):-N<R,Bs=[N].
to_base(N,R,K,[B|Bs]):-N>=R,
  B is N mod R, N1 is N//R,K1 is K+1,
  to_base(N1,R,K1,Bs).

% conversion from list of digits in given base
from_base(_Base,[],0).
from_base(Base,[X|Xs],N):-from_base(Base,Xs,R),N is X+R*Base.

% conversion to list of bits, right to left 
to_rbits(N,Bs):-to_base(2,N,Bs).

% conversion from list of bits, right to left 
from_rbits(Bs,N):-from_base(2,Bs,N).

% counting how many bits a number needs     
bitcount(N,K):-N=<1,K=1.
bitcount(N,K):-N>1,N1 is N>>1,bitcount(N1,K1),K is K1+1.

% finds the larges bitcount for a list
max_bitcount(Nss,L):-maplist(bitcount,Nss,Ls),max_list(Ls,L).  
  
% pads up to maxbits, if needed
to_maxbits(Maxbits,N,Rs):-
  to_base(2,N,Bs),length(Bs,L),ML is Maxbits-L,
  ndup(ML,0,Zs),append(Bs,Zs,Rs).
\end{code}
\end{document}